\documentclass[aps,amsmath,amssymb,amsfonts,twocolumn,prl,showpacs]{revtex4-1}
\usepackage{graphicx}
\usepackage{bm}
\usepackage{color}
\usepackage{amsmath,amsthm,amssymb}
\usepackage{subfigure}

\usepackage{epstopdf}

\newcommand{\mbf}{\mathbf}
\newcommand{\mrm}{\mathrm}

\begin{document}
\author{Micha{\l} Krych}
\affiliation{Faculty of Physics, University of Warsaw, Pasteura 5, 02-093 Warsaw, Poland}
\author{Zbigniew Idziaszek}
\affiliation{Faculty of Physics, University of Warsaw, Pasteura 5, 02-093 Warsaw, Poland}
\title{Description of ion motion in a Paul trap immersed in a cold atomic gas}

\begin{abstract}
We investigate the problem of a single ion in a radio-frequency trap and immersed in an ultracold Bose gas either in a condensed or a non-condensed phase. We develop master equation formalism describing the sympathetic cooling and we determine the cooling rates of ions. We show that cold atomic reservoir modifies the stability diagram of the ion in the Paul trap creating the regions where the ion is either cooled or heated due to the energy quanta exchanged with the time-dependent potential.
\end{abstract}


\maketitle

\section{I. Introduction}
Hybrid systems combining cold atomic gases with single ions or ionic crystals attract an increasing attention \cite{Grier2009,Zipkes2010,Schmid2010,Rellergert2011,Ravi2011,Hall2011,Hall2012,Sivarajah2012,Denschlag2013,Denschlag2013a}. They have been proposed for implementation of quantum gates \cite{Idziaszek2007,Doerk2010,Nguyen2012}, realization of new mesoscopic quantum states  \cite{Cote2002,Massignan2005}, probing quantum gases \cite{Sherkunov2009}, studying controlled chemical reactions at low temperatures \cite{Rellergert2011,Hall2011,Hall2012,Hall2013} or emulating some well-known condensed-matter physics phenomena \cite{Gerritsma2012,Bissbort2013}.
The theoretical framework to describe atom-ion collisions in the quantum regime has been developed \cite{Cote2000,Bodo2008,Idziaszek2009,Gao2010,Idziaszek2011,Gao2011,Gao2013}, however the ab-initio potentials
{\cite{Makarov2003,Knecht2010,Krych2011,Hall2013,Hall2013a, Tomza2014}} are not known with accuracy sufficient for precise determination of scattering lengths. Their values can be measured in experiments, for instance by applying technique of Feshbach resonances \cite{Idziaszek2009}, provided the ions immersed in cold atomic gas are cooled down to the quantum regime where scattering takes place only in the lowest partial waves. Such low temperatures can be reached, for instance, via sympathetic cooling of ions in contact with cold atomic reservoir \cite{Goodman2012,Hudson2009,Ravi2011}. This method, however, suffers both due to some technical issues (e.g. excess micromotion \cite{Denschlag2013}), or due to some fundamental limitations resulting from the ion dynamics in the time-dependent Paul trap \cite{Dehmelt1968,Ravi2011,Vuletic2012,Goodman2012}. Apart from the Paul traps there are first successful experimental attempts on optical ion trapping {\cite{Huber2014,Schneider2010}}. So far the problem of atom-ion sympathetic cooling has been studied only in the classical regime \cite{DeVoe2010,Vuletic2012,Chen2013}. The purpose of this paper is to provide a consistent framework to describe the process of sympathetic cooling in the quantum regime and to study its limitations for experimental systems where a single ion is immersed in a Bose-Einstein condensate \cite{Zipkes2010,Schmid2010}.

{The paper is organized as follows.} {In Sec. II. we {introduce} the master equation formalism and {discuss} two reservoirs: an non-condensed ultracold gas and a {Bose-Einstein} condensate. A regularization of an atom-ion interaction potential is presented in Sec. III. Next, in Sec. IV. we derive the master equation. After that, in Sec. V. we discuss the evolution of the position operator in the time-dependent Paul trap. In Sec. VI. we present the equations of motion of the ion in contact with the cold reservoir. Finally, in Sec. VII. we discuss the ion cooling rates {and stability regimes} for different experimental parameters.}

\section{II. Master equation formalism} We describe the system treating the atomic gas as a reservoir, and deriving an effective equation for the dynamics of the ion. The total Hamiltonian consists of the following parts: $\hat{H}=\hat{H}_S+\hat{H}_R+\hat{H}_{RS}$, where $\hat{H}_S$ is the Hamiltonian of the ion \cite{leibfried}
\begin{equation}
\begin{split}
\label{Ham1}
\hat{H}_S & =\frac{\hat{\mbf{p}}^2}{2M} +  \frac{M}2  \sum_j \omega_j^2(t) \hat{r}_j^2,
\end{split}
\end{equation}
where the time-dependent trapping frequency consists of the static and dynamic parts: $\omega_j^2(t) = \Omega^2\frac14\left( a_j+ 2 q_j \cos \left( \Omega t \right)\right)$ and {$j=x,y,z$ is a spatial direction.}
Here, $\hat{\mbf{p}} = (\hat{p}_x,\hat{p}_y,\hat{p}_z)$ and $\hat{\mbf{r}} = (\hat{r}_x,\hat{r}_y,\hat{r}_z)$ denote the momentum and position operators, respectively, $M$ is the ion mass and $\Omega$ is the radio frequency of the dynamic part. A homogeneous gas of atoms is described by
{\begin{equation}\label{HamR}
\begin{split}
\hat{H}_R&= \int \!\! d^3 r_a \hat{\Psi}^\dagger(\mbf{r}_a) \left(\frac{\hat{\mbf{p}}_a^2}{2 m} +
\frac{g}{2} \hat{\Psi}^\dagger(\mbf{r}_a) \hat{\Psi}(\mbf{r}_a) \right) \hat{\Psi}(\mbf{r}_a)\\
&=\sum_{\mbf{k}}\hbar \omega_{\mbf{k}} \hat{a}^\dagger_\mbf{k} \hat{a}_\mbf{k},
\end{split}
\end{equation}}
{where $\hat{\Psi}(\mbf{r}_{\hat{a}})=\sum_{\mbf{k}}e^{i \mbf{k} \mbf{r}} {\hat{a}}_\mbf{k}/\sqrt{L^3}$ is the field operator, $L$ is the size of the quantization box, $a_\mbf{k}$ is the annihilation operator for mode $\mbf{k}$, $\hbar\omega_k$ is energy of this mode, $\hat{\mbf{p}}_a$ is the {atomic} momentum operator, $m$ is the atomic mass, and $g = 4 \pi \hbar^2 a /m $ is the interaction constant with $a$ denoting the $s$-wave scattering length. The ion-atom interaction is given by}
\begin{equation}\label{HamRS}
\hat{H}_{RS} = \int d^3 r_a \hat{\Psi}^\dagger(\mbf{r}_a) V(\mbf{r}_a - \hat{\mbf{r}}) \hat{\Psi}(\mbf{r}_a),
\end{equation}
where $V(\mbf{r})$ is the atom-ion interaction potential. In our approach we treat the atomic gas in the second quantization formalism, while the ion is described by position and momentum  operators.


For an ion immersed in a Bose-Einstein condensate (BEC) we describe the reservoir in a Bogoliubov approximation
\begin{equation}\label{Bogoliubov}
\hat{H}_R=E_0+\sum_{\mbf{q}}\varepsilon(\mbf{q})\hat{b}^\dagger_\mbf{q}\hat{b}_\mbf{q}
\end{equation}
where $\hat{b}^\dagger_\mbf{q}$ and $\hat{b}_\mbf{q}$ are the creation and annihilation operators for Bogoliubov excitations with momentum $\hbar \mbf{q}$ and energy $\varepsilon(\mbf{q})$ and $E_0$ is the ground state energy of the superfluid. {Even in the ground state of the Paul trap the speed of motion of the ion is typically much larger than the speed of sound in the condensate, and ion couples only to the particle part of the Bogoliubov spectrum $\varepsilon(\mbf{q})\approx \hbar^2 q^2/(2m)$.} The ion and the superfluid are coupled by the density-density interaction \cite{Pitaevskii,DaleyZoller2004}
\begin{equation}\label{oddzialywanieBEC}
\begin{split}
\hat{H}_{RS}&=\int d^3 r d^3 r' V(\mbf{r}-\mbf{r}')\delta\hat{\rho}(\mbf{r})\delta\hat{\rho}_{ion}(\mbf{r}')\\
&=\int d^3 r V(\mbf{r}-\mbf{\hat{r}})\delta\hat{\rho}(\mbf{r})
\end{split}
\end{equation}
where $\delta\hat{\rho}(\mbf{r})=\hat{\mbf{\Psi}}^\dagger \hat{\mbf{\Psi}}-\rho_0$ and  $\hat{\mbf{\Psi}}=\sqrt{\rho_0}+\delta\hat{\mbf{\Psi}}$ is the quantized field operator for the superfluid and $\rho_0$ is the condensate density, $\delta\hat{\rho}_{ion}(\mbf{r}')=\delta^3 (\mbf{r}'-\hat{\mbf{r}})$ is the density operator of the position of the ion. The field operator for the excitations of the superfluid is given by \cite{DaleyZoller2004} $\delta\hat{\Psi}=L^{-3/2}\sum_\mbf{q}\left(u_\mbf{q} \hat{b}_\mbf{q} exp(i\mbf{q r})+v_\mbf{q} \hat{b}_\mbf{q}^\dagger exp(-i\mbf{q r})\right)$, where $L$ is the size of a box, and we assume the periodic boundary conditions.



\section{III. Regularized potential}
\label{App:RegulPot}
\begin{figure}\includegraphics[width=0.6\linewidth,clip]{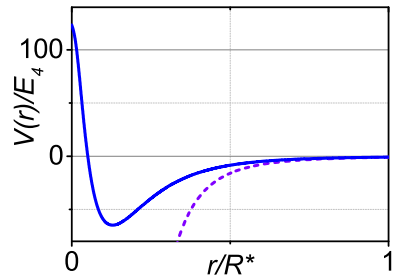}
\caption{\label{Fig:potencjal}
(Color online). A regularized potential (solid line) and a pure $-C_4/r^4$ (dashed line).
}
\end{figure}

Long range interaction between ion and atom is described by the polarization potential $-C_4/r^4$, but at short distances this potential is singular and it needs to be regularized. We introduce a regularized version of the polarization potential
\begin{equation}
\label{potential}
V(r)=-C_4\frac{r^2-c^2}{r^2+c^2}\frac{1}{(b^2+r^2)^2}
\end{equation}
that mimics at large distances the behavior of $-C_4/r^4$ tail. For small separations between particles it supports a single minimum {(cf. Fig.~\ref{Fig:potencjal})}. The short-range repulsive part is finite, which simplifies numerical calculations. We choose $b$ and $c$ parameters in such a way, that scattering amplitude calculated in the first-order Born approximation
\begin{equation}
\begin{split}
\label{fq}
f(q)&= \frac{\pi (R^{\star})^2}{4b(b^2-c^2)^2 q}\left(-4b c^2 e^{-cq}\right.\\
&\left.+e^{-bq}(4b c^2+(b^2-c^2)(b^2+c^2)q)\right)
\end{split}
\end{equation}
is equal to the exact scattering amplitude of the potential at zero energy. Here, $q=|\mbf{k}-\mbf{k}^\prime|$ is the magnitude of momentum transfer.
In this way our description within the master equation formalism \cite{Carmichael}, treating system-reservoir interactions in the Born approximation will be accurate for a single collision in the ultracold regime.

With the polarization potential one can associate a characteristic length $R^\ast = \left( 2 \mu C_4/\hbar^2\right)^{1/2}$ and a characteristic energy $E^\ast = \hbar^2/\left[2 \mu (R^\ast)^2\right]$.
We choose $b$, $c$ parameters in such a way, that a scattering amplitude calculated in the first order Born approximation
is equal to the exact scattering amplitude of the potential at zero energy.
For example for $b=0.0781R^{\star}$ and $c=0.2239R^{\star}$ potential \eqref{potential} supports a single bound state, its scattering length is equal $a_{sc}=R^{\star}$ and the
zero-energy scattering amplitudes calculated exactly and from Eq.~\eqref{fq} are equal to $-R^{\star}$.

\section{IV. Master equation}\label{App:Srednie}

The interaction Hamiltonian of the ion with the ultracold gas (Eq.~\eqref{HamRS}) can be rewritten explicitly as
\begin{equation}\label{oddzialywaniegazrozwiniete}
\begin{split}
\hat{H}_{RS}&=\sum_{\mbf{k},\mbf{k}'}\hat{a}_\mbf{k'}^\dagger \hat{a}_{\mbf{k}}  L^{-3}\int d^3 r_a e^{-i(\mbf{k}'-\mbf{k})\mbf{r}_a} V(\mbf{r}_a-\hat{\mbf{r}}_j)\\
&=\sum_{\mbf{k},\mbf{k}'}\hat{a}_\mbf{k'}^\dagger \hat{a}_{\mbf{k}} e^{-i(\mbf{k}'-\mbf{k})\hat{\mbf{r}}_j}  L^{-3}\int d^3 r e^{-i(\mbf{k}'-\mbf{k})\mbf{r}} V(\mbf{r}),
\end{split}
\end{equation}
where $V(\mbf{r})$ denotes the interaction potential. It is easy to {separate ion and reservoir operators in a general form used in derivation of the master equation \cite{Carmichael}}
\begin{equation}
\hat{H}_{RS}=\hbar \sum_{\mbf{k},\mbf{k}'} \hat{s}_{\mbf{k}\mbf{k}'} \hat{\Gamma}_{\mbf{k}\mbf{k}'},
\end{equation}
where the ion part reads
\begin{equation}
\hat{s}_{\mbf{k}\mbf{k}'}=e^{-i(\mbf{k}'-\mbf{k})\hat{\mbf{r}}} c_{\mbf{k},\mbf{k}'}/\hbar,
\end{equation}
and $c_{kk'}$ is a Fourier transform of the interaction potential
\begin{equation}
c_{\mbf{k},\mbf{k}'}= L^{-3} \int d^3 r e^{i(\mbf{k}-\mbf{k}')\hat{\mbf{r}}} V(\mbf{r}).
\end{equation}
Gas operators can be written down as
\begin{equation}
\hat{\Gamma}_{\mbf{k}\mbf{k}'}=\hat{a}_\mbf{k}^\dagger \hat{a}_{\mbf{k}'}.
\end{equation}
General textbook form of a master equation {for a reduced density matrix} (page 8 in \cite{Carmichael}) after the Markov approximation reads
\begin{equation}\label{masterksiazka}
\begin{split}
\dot{\hat{\tilde{\rho}}}=\!\!\!\!\sum_{\mbf{k},\mbf{k}',\mbf{l},\mbf{l}'}\!\int_0^t \!\!\!\! & dt'  \!\left( [\hat{\tilde{s}}_{\mbf{l}\mbf{l}'}(t')\hat{\tilde{\rho}}(t),\hat{\tilde{s}}_{\mbf{k}\mbf{k}'}(t)]\langle\hat{\tilde{\Gamma}}_{\mbf{k}\mbf{k}'}(t)\hat{\tilde{\Gamma}}_{\mbf{l}\mbf{l}'}(t')\rangle_{R_0} \right.\\
&\left.+ [\hat{\tilde{s}}_{\mbf{k}\mbf{k}'}(t),\hat{\tilde{\rho}}(t)\hat{\tilde{s}}_{\mbf{l}\mbf{l}'}(t')]\langle\hat{\tilde{\Gamma}}_{\mbf{l}\mbf{l}'}(t')\hat{\tilde{\Gamma}}_{\mbf{k}\mbf{k}'}(t)\rangle_{R_0}\right).
\end{split}
\end{equation}
Here $\langle\dots\rangle_{R_0}$ is a trace over the reservoir density matrix $R_0$, tilded operators denote an interaction picture with respect to the noninteracting system (ion in a Paul trap Eq.~\eqref{Ham1}) and a reservoir (noninteracting gas Eq.~\eqref{HamR})
\begin{equation}
\hat{\tilde{X}}(t)=U^\dagger (0,t) e^{(i/\hbar)\hat{H}_R t} \hat{X} e^{-(i/\hbar)\hat{H}_R t}U (0,t)
\end{equation}
where $U(t_1,t_2)={\cal{T}}\exp(-\frac{i}{\hbar}\int_{t_1}^{t_2}d\tau \hat{H}_S(\tau))$ is the time-ordered evolution operator of an ion immersed in a Paul trap. Let us consider the free evolution of the gas operators
\begin{equation}
\hat{\tilde{\Gamma}}_{\mbf{k}\mbf{k}'}(t)=\hat{\tilde{a}}_\mbf{k}^{\dagger}(t)\hat{\tilde{a}}_{\mbf{k}'}(t)=\hat{a}_\mbf{k}^{\dagger}\hat{a}_{\mbf{k}'} e^{i(\omega_\mbf{k}-\omega_{\mbf{k}'})t}.
\end{equation}
We denote the mean occupation number of the atomic modes as
\begin{equation}
\bar{n}_\mbf{k}\equiv \langle \hat{a}_\mbf{k}^{\dagger}\hat{a}_{\mbf{k}}\rangle_{R_0}
\end{equation}
where $\bar{n}_\mbf{k}=1/(\exp(\hbar^2 \mbf{k}^2/(2m k_B T)-\mu)-1)$ and $\mu$ denotes the chemical potential, $k_B$ is the Boltzmann constant nad $T$ is a temperature.  Using the commutation relations for creation and annihilation operators we are able to eliminate the reservoir degrees of freedom and rewrite the master equation with an explicit form of gas correlation functions
\begin{equation}\label{masterkolejne}
\begin{split}
\dot{\hat{\tilde{\rho}}}=&\sum_{\mbf{k},\mbf{k}'}\int_0^t dt'\\
&\left([\hat{\tilde{s}}_{\mbf{k}' \mbf{k}}(t')\hat{\tilde{\rho}}(t),\hat{\tilde{s}}_{\mbf{k}\mbf{k}'}(t)] e^{i(t-t')(\omega_\mbf{k}-\omega_{\mbf{k}'})} \bar{n}_{\mbf{k}}(\bar{n}_\mbf{k'}+1)\right. \\
&+\left.[\hat{\tilde{s}}_{\mbf{k}\mbf{k}'}(t),\hat{\tilde{\rho}}(t)\hat{\tilde{s}}_{\mbf{k}' \mbf{k}}(t')] e^{i(t-t')(\omega_\mbf{k}-\omega_{\mbf{k}'})}  \bar{n}_{\mbf{k'}}(\bar{n}_\mbf{k}+1)\right).
\end{split}
\end{equation}
In the next step we {transform the master equation back from the interaction picture}
\begin{equation}
\dot{\hat{\rho}}=\frac{1}{i\hbar}[\hat{H}_S,\hat{\rho}]+U(0,t) \dot{\hat{\tilde{\rho}}} U^\dagger(0,t).
\end{equation}
{ After changing the order of terms in commutators and of the summation indices in the last line of {\eqref{masterkolejne}} and substituting $t-t'\equiv \tau$, the master equation reads
\begin{equation}\label{masterkolejnekolejne}
\begin{split}
\dot{\hat{\rho}}&=\frac{1}{i\hbar}[\hat{H}_S,\hat{\rho}]-\sum_{\mbf{k},\mbf{k}'} \bar{n}_\mbf{k}(\bar{n}_{\mbf{k}'}+1)\\
&\times \int_0^t  d\tau \left( e^{i\tau(\omega_\mbf{k}-\omega_{\mbf{k}'})} \left[{\hat{s}}_{\mbf{k}\mbf{k}'},{\hat{s}}_{\mbf{k}' \mbf{k}}(t,-\tau){\hat{\rho}}\right]\right.\\
&+\left.e^{-i\tau(\omega_\mbf{k}-\omega_{\mbf{k}'})} \left[{\hat{\rho}} {\hat{s}}_{\mbf{k} \mbf{k}'}(t,-\tau),{\hat{s}}_{\mbf{k}' \mbf{k}}\right]\right).
\end{split}
\end{equation}
For any operator $\hat{x}$ we define
\begin{equation}
\begin{split}
\hat{x}(t,-\tau)&\equiv U(0,t)U^\dagger(0,t-\tau)\hat{x}U(0,t-\tau)U^\dagger(0,t)\\
&\equiv U(0,t)\hat{x}(t-\tau)U^\dagger(0,t)
\end{split}
\end{equation}
 In case of a time independent Hamiltonian this would reduce to $\hat{x}(t,-\tau)=U^\dagger(0,-\tau)\hat{x}U(0,-\tau)$. However, we note that in a Paul trap the evolution cannot be reduced to the single evolution operator with the time difference as in energy-conserving system, since $U(0,t)U^\dagger(0,t-\tau) \neq U^\dagger(0,-\tau)$.
  The integration over $\tau$ in the master equation is dominated by short timescales, because correlation functions in a large reservoir vanish quickly in time.}
  In this way we can extend the integration limit up to the infinity, as is usually done in the derivation of the master equation \cite{Carmichael}. {Substituting an explicit form of operators $\hat{s}_{\mbf{k}\mbf{k}'}$ into \eqref{masterkolejnekolejne} yields}
\begin{equation}\label{masterexp}
\begin{split}
\dot{\hat{\rho}}&=\frac{1}{i\hbar}[\hat{H}_S,\hat{\rho}]-\sum_{\mbf{k},\mbf{k}'}\bar{n}_{\mbf{k}}(\bar{n}_{\mbf{k}'}+1) c_{\mbf{k},\mbf{k}'} c_{\mbf{k}',\mbf{k}}/\hbar^2 \\
 &\times \int_0^{\infty}\!\!\! d\tau\left( e^{i\tau (\omega_\mbf{k}-\omega_{\mbf{k}'})} \left[e^{i(\mbf{k}-\mbf{k}')\hat{\mbf{r}}}, e^{-i(\mbf{k}-\mbf{k}')\hat{\mbf{r}}(t,-\tau)}{\hat{\rho}}\right]\right.\\
&+\left. e^{-i\tau (\omega_\mbf{k}-\omega_{\mbf{k}'})} \left[{\rho}e^{i(\mbf{k}-\mbf{k}')\hat{\mbf{r}}(t,-\tau)}, e^{-i(\mbf{k}-\mbf{k}')\hat{\mbf{r}}}\right]\right).
\end{split}
\end{equation}
 {In case of a non-condensed buffer gas the dynamics of the reduced density operator $\hat{\rho}$ obtained by tracing over the reservoir modes, derived within Born and Markov approximations \cite{Gardiner,Carmichael} \eqref{masterexp} can be straightforwardly rewritten in a compact form}
\begin{equation}\label{MasterGas}
\dot{\hat{\rho}}(t)= \frac{1}{i\hbar}\left[\hat{H}_S(t),\hat{\rho}\right]-\sum_{\mbf{k},\mbf{k}'} \Omega_{\mbf{k},\mbf{k}'}^2
\left\{
\left[
\hat{Z}_{\mbf{k},\mbf{k}'},\hat{W}_{\mbf{k},\mbf{k}'}(t) \hat{\rho} \right] + H. c.
\right\}
\end{equation}
{where $\mbf{k}$ and $\mbf{k}^\prime$ are the quantized wave vectors of atoms in a box {of size $L^3$}, $\Omega_{\mbf{k},\mbf{k}'}^2 = \bar{n}_\mbf{k} (\bar{n}_{\mbf{k}'}+1) |c_{\mbf{k},\mbf{k}'}|^2/\hbar^2$.
Furthermore, $\hat{Z}_{\mbf{k},\mbf{k}'} = e^{i(\mbf{k}-\mbf{k}')\hat{\mbf{r}}}$, and
$\hat{W}_{\mbf{k},\mbf{k}'}(t) = \int_{0}^{\infty} d\tau
e^{i\tau (\omega_{\mbf{k}}-\omega_\mbf{k}')}
e^{i(\mbf{k}'-\mbf{k})\hat{\mbf{r}}(t,-\tau)}$.}

 For the Bose condensed reservoir {in order to derive the master equation we use the interaction Hamiltonian given by Eq.~\eqref{oddzialywanieBEC} instead of Hamiltonian \eqref{oddzialywaniegazrozwiniete} and we describe the reservoir in the Bogoliubov approximation (Eq.~\eqref{Bogoliubov}). All the following steps of the derivation are analogous to the non-condensed case: (i) {we start with} the general textbook form of the master equation for a reduced density matrix after the Markov approximation (Eq.~\eqref{masterksiazka}), (ii) perform the trace operation with respect to the BEC degrees of freedom, (iii) {transform back} from the interaction picture, (iv) extend the time integration up to the infinity. {Therefore in case of the BEC reservoir the master equation reads}
\begin{equation}\label{MasterBose}
\begin{split}
\dot{\hat{\rho}}(t)&= \frac{1}{i\hbar}\left[\hat{H}_S(t),\hat{\rho}\right]-\sum_{\mbf{q}} \Omega_{\mbf{q}}^2 \left\{\bar{n}_\mbf{q}\left[\hat{Z}_{0,\mbf{q}},\hat{W}_{0,\mbf{q}}(t) \hat{\rho} \right]\right.\\
&\left.+(1+\bar{n}_\mbf{q})\left[
\hat{\rho}\hat{W}_{0,\mbf{q}}(t),\hat{Z}_{0,\mbf{q}}  \right] + H. c.
\right\}
\end{split}
\end{equation}
where $\Omega_{\mbf{q}}^2=\rho_0 L^3|c_{0,\mbf{q}}|^2/\hbar^2$ {and $\bar{n}_\mbf{q}=1/(\exp(\hbar^2 \mbf{q}^2/(2m k_B T))-1)$ denotes the mean occupation number of the Bogoliubov quasiparticles}. {We note that the master equation for a BEC (Eq.~\eqref{MasterBose}) is equivalent to the low temperature limit of the master equation for the non-condensed gas (Eq.~\eqref{MasterGas}). This is because of the fact that for typical parameters the ion couples only to the particle region of the Bogoliubov excitations.}

\section{V. Time dependence of the position operator}

In order to determine the coefficients of master equations \eqref{MasterGas} and \eqref{MasterBose} we have to find the {evolution of the position operator $\hat{\mbf{r}}(t,-\tau)$ of the ion in a Paul trap in the absence of the ultracold gas.} We start from Heisenberg equations of motion for position  $\dot{\hat{r}}_j=\hat{p}_j/M$ and momentum $\dot{\hat{p}}_j=-M \omega_j^2(t) \hat{r}_j$ derived with help of Hamiltonian from Eq.~\eqref{Ham1}, where $j=x,y,x$ denotes the direction in space. Their combination leads to the second order differential equation
\begin{equation}
\label{heisenberg}
\ddot{\hat{r}}_j+\omega_j^2(t) \hat{r}_j=0,
\end{equation}
where the time-dependent trapping frequency consists of the static and dynamic parts: $\omega_j^2(t) = \Omega^2\frac14\left( a_j+ 2 q_j \cos \left( \Omega t \right)\right)$, as was defined before. Above equation has two linearly independent $\cal{C}$-number solutions  $u_j(t)$ and $u_j(-t)$, where  $u_j(t)=e^{i(\beta_j/2) \Omega t} \sum_{n=-\infty}^{\infty}C^j_{n} e^{in\Omega t}$, $u_j(0)=1$ and  $C^j_n$ are expansion coefficients of Mathieu functions describing the time evolution of a single ion in a Paul trap and $(\beta_j/2) \Omega$ is an effective secular frequency \cite{leibfried}. Eq.~\eqref{heisenberg} does not have the first order derivative, so {the Wronskian  of $u_j(t)$ and $u_j(-t)$ is constant in time}
\begin{equation}\label{wr0}
\begin{split}
W\left(u_j(t),u_j(-t)\right)&=u_j(-t)\dot{u}_j(t)-u_j(t)\dot{u}_j(-t)\\
                 &=2 i \nu_j=\mrm{const},
\end{split}
\end{equation}
where $\nu_j=\dot{u}_j(0)/i=\Omega \sum_{n=-\infty}^{\infty} C^j_{n} (\beta_j/2+n)$ {is called a reference harmonic oscillator frequency. One can also define two other Wronskians between position operator and $u_j(t)$, which is proportional to}
\begin{equation}\label{wr1}
\begin{split}
\hat{c}_{j,1}(t)&=i\sqrt{M/(2\hbar \nu_j)}\times  W\left({\hat{r}}_j(t),u_j(t)\right)\\
&=i\sqrt{M/(2\hbar \nu_j)}\left(u_j(t)\dot{\hat{r}}_j(t)-\dot{u}_j(t)\hat{r}_j(t)\right)\\
&=\hat{c}_{j,1}(0)=1/\sqrt{2 M \hbar \nu_j}(M \nu_j \hat{r}_j(0)+i \hat{p}_j(0))
\end{split}
\end{equation}
{or between position operator and  $u_j(-t)$, which is proportional to}
\begin{equation}\label{wr2}
\begin{split}
\hat{c}_{j,2}(t)&=-i\sqrt{M/(2\hbar \nu_j)}\times  W\left({\hat{r}}_j(t),u_j(-t)\right)\\
&=-i\sqrt{M/(2\hbar \nu_j)}\left(u_j(-t)\dot{\hat{r}}_j(t)-\dot{u}_j(-t)\hat{r}_j(t)\right)\\
&=\hat{c}_{j,2}(0)=1/\sqrt{2 M \hbar \nu_j}(M \nu_j \hat{r}_j(0)-i \hat{p}_j(0))
\end{split}
\end{equation}
 {Since Wronskians of Eq.~\eqref{heisenberg} must be constant in time, operators $\hat{c}_{j,1}(t)$ and $\hat{c}_{j,2}(t)$ are constant.} For $\beta_j$ real one can show that they are equivalent to creation and annihilation operator (respectively) of the reference harmonic oscillator of frequency $\nu_j$ \cite{leibfried}. Both of the above equations connect $\hat{r}_j(t)=U^\dagger(0,t)\hat{r}_j(0)U(0,t)$ and  $\hat{p}_j(t)=U^\dagger(0,t)\hat{p}_j(0)U(0,t)$ with their values for $t=0$ ($\hat{r}_j(0)\equiv\hat{r}_j$ and $\hat{p}_j(0)\equiv\hat{p}_j$).
Multiplying Eqs.~\eqref{wr1}, \eqref{wr2} by $u_j(-t)$ and $u_j(t)$, respectively, adding the first one to the second one and using Eq.~\eqref{wr0} we can express $\hat{r}_j(t)$ as a function of $\hat{r}_j(0)$ and $\hat{p}_j(0)$.
\begin{equation}\label{wr3}
\begin{split}
\hat{r}_j(t)&=\frac{\hat{r}_j(0)}{2}(u_j(t)+u_j(-t))+\frac{\hat{p}_j(0)}{2 i M \nu_j}(u_j(t)-u_j(-t))
\end{split}
\end{equation}
Basing on Heisenberg equations of motion $\hat{p}_j=M \dot{\hat{r}}_j$ {we can express} $\hat{p}_j(t)$ also as a function of $\hat{r}_j(0)$ and $\hat{p}_j(0)$.
\begin{equation}\label{wr4}
\begin{split}
\hat{p}_j(t)&=\frac{M\hat{r}_j(0)}{2}(\dot{u}_j(t)+\dot{u}_j(-t)+\frac{\hat{p}_j(0)}{2 i \nu_j}(\dot{u}_j(t)-\dot{u}_j(-t)))
\end{split}
\end{equation}
{Starting from Eqs.~\eqref{wr3} and \eqref{wr4} it} is straightforward to derive the following result
\begin{equation}
\begin{split}\label{r0}
\hat{r}_j(0)&=\frac{\hat{r}_j(t)}{2i\nu_j}(\dot{u}_j(t)-\dot{u}_j(-t))-\frac{\hat{p}_j(t)}{2 i M \nu_j}(u_j(t)-u_j(-t))
\end{split}
\end{equation}
and
\begin{equation}\label{p0}
\begin{split}
\hat{p}_j(0)&=-\frac{M\hat{r}_j(t)}{2}(\dot{u}_j(t)+\dot{u}_j(-t))+\frac{\hat{p}_j(t)}{2}(u_j(t)+u_j(-t)).
\end{split}
\end{equation}
Now we are able to calculate{
\begin{widetext}
\begin{equation}
\begin{split}
\hat{r}_j(t,-\tau)&\equiv  U(0,t) U^\dagger(0,t-\tau) \hat{r}_j U(0,t-\tau) U^\dagger(0,t) = U(0,t)\hat{r}_j(t-\tau) U^\dagger(0,t)\\
&=U^\dagger  (t,0)\left(\frac{\hat{r}_j(0)}{2}(u_j(t-\tau)+u_j(-t+\tau))+\frac{\hat{p}_j(0)}{2 i M \nu_j}(u_j(t-\tau)-u_j(-t+\tau))\right)U(t,0)
\end{split}
\end{equation}
\end{widetext}}
where we have used the identity $U^\dagger(0,t)\equiv U(t,0)$. {With help of} Eqs.~\eqref{r0} and \eqref{p0} we perform the time evolution from $t$ to $0$.
With an explicit form of $u_j(\pm t)$ this can be arranged in a compact form
\begin{equation}\label{rodtau}
\hat{r}(t,-\tau)=\sum_{n,m}\!C_n C_m\!\!\left[\hat{r}  \left(\frac{\!\beta}{2}\!+\!m\!\right)\!\frac{\Omega}{\nu} \cos I_{nm}^\tau\!-\!\frac{\hat{p}}{\nu M} \sin I_{nm}^\tau\right]
\end{equation}
where we have omitted $j$ index for simplicity, $I_{nm}^\tau\equiv \Omega\left(\left(\frac{\beta}{2}+n\right)\tau-(n-m)t\right)$.

\section{VI. Equations of motion}\label{App:Srednie}

We expand exponential terms of master equation (Eq.~\eqref{masterexp}) in the small Lamb-Dicke parameter $\zeta=a_{i}/\lambda_T$ up to the second-order terms, where $a_{i}=\sqrt{\hbar/(M\nu)}$ is a length scale of the secular potential with a reference oscillator frequency (which is {comparable to the size of the ion wavefunction}) and $\lambda_T=\sqrt{2\pi \hbar^2/(m k_B T)}$ is de Broglie wavelength, with $T$ denoting the temperature of the reservoir (a typical change of the atomic momenta $(\mbf{k}-\mbf{k'})$ during a single atom-ion collision is of the order of $\lambda_T^{-1}$). For example
\begin{equation}
\begin{split}
&\left[e^{i(\mbf{k}-\mbf{k'})\hat{\mbf{r}}}, e^{-i(\mbf{k}-\mbf{k'})\hat{\mbf{r}}(t,-\tau)}{\rho}\right]=[i(\mbf{k}-\mbf{k'})\hat{\mbf{r}},\rho]+\\
&+\sum_j (k_j-k'_j)^2(\frac12[\rho, \hat{r}_j^2]+[\hat{r}_j, \hat{r}_j(t,-\tau)\rho])+\dots
\end{split}
\end{equation}
where $j=x,y,z$. We {note that the terms with odd powers of $(k_j-k_j')$ vanish due to the symmetric} summation in the master equation. Every odd term of the expansion is antisymmetric in $(k_j-k_j')$, so the first neglected term is of the fourth order in a small Lamb-Dicke parameter $\zeta$. The master equation exact up to the third order in $\zeta$ is given by
\begin{equation}
\begin{split}
\dot{\hat{\rho}}=&\frac{1}{i\hbar}[\hat{H}_S,\hat{\rho}]-\sum_j\sum_{k_j,k_j'}\frac{c_{\mbf{k},\mbf{k}'}c_{\mbf{k}', \mbf{k}}}{2\hbar^2} (k_j-k_j')^2 \bar{n}_{\mbf{k}}(\bar{n}_{\mbf{k}'}+1)\\&\times\int_0^{\infty} d\tau
 \left(e^{i\tau (\omega_{\mbf{k}}-\omega_{\mbf{k}'})} \left([\hat{\rho},\hat{r}_j^2]+2[\hat{r}_j, \hat{r}_j(t,-\tau)\hat{\rho}]\right)\right.\\
&-\left. e^{-i\tau (\omega_{\mbf{k}}-\omega_{\mbf{k}'})} \left([\hat{r}_j^2,\hat{\rho}]+2[\hat{r}_j,\hat{\rho} \hat{r}_j(t,-\tau)]\right)\right).
\end{split}
\label{2rzad}
\end{equation}
{We note that in this approximation spatial directions are not coupled.}

In order to derive equations of motion {for expectation values of position and momentum operators} we multiply master equation for ultracold gas or BEC, respectively, by $\hat{r}$ or $\hat{p}$ operators and perform tracing over ion degrees of freedom.
In this way we obtain
\begin{equation}
\label{liniowej}
\begin{split}
\dot{\overline{r}}_j&=\overline{p}_j/M\\
\dot{\overline{p}}_j&=- M\omega_j^2(t) \overline{r}_j +K^j(t)\overline{r}_j- 2 G^j_{\eta,\delta}(t)\overline{p}_j,
\end{split}
\end{equation}
{where in general, expectation values are defined as follows
 \begin{equation}
 {\overline{x}}\equiv\mrm{tr}\left\{\hat{x} {\rho}\right\}
 \end{equation}
 \begin{equation}
 \dot{\overline{x}}\equiv\mrm{tr}\left\{\hat{x} \dot{\rho}\right\}
 \end{equation}}

Similar procedure may be applied to equations involving expectation values of operators quadratic in position and momentum
\begin{align}
\label{kwadratowej}
(\dot{\overline{r_j p_j}}+\dot{\overline{p_j r_j}})=&-2M\omega_j^2(t) \overline{r_j^2}+2\overline{p_j^2}/M+2K_j(t)\overline{r_j^2} \nonumber\\
&-2G^j_{\eta,\delta}(t)(\overline{r_j p_j}+\overline{p_j r_j})+2\hbar G^j_{\gamma,-\mu}(t),\nonumber\\
\dot{\overline{r_j^2}}=&\frac{1}{M} (\overline{r_j p_j}+\overline{p_j r_j}),\\
\dot{\overline{p_j^2}}=&-M\omega_j^2(t) (\overline{r_j p_j}+\overline{p_j r_j})-4 G^j_{\eta,\delta}(t)\overline{p_j^2}\nonumber\\
&+K^j(t) (\overline{r_j p_j}+\overline{p_j r_j})+2\hbar D^j_{\mu,\gamma}(t).\nonumber
\end{align}
Coefficients in Eqs.~\eqref{liniowej} and \eqref{kwadratowej} have the following form
\begin{align}
G^j_{\eta,\delta}(t) = & \sum_{n,m}\frac{C_n^j C_m^j}{\nu_j}\frac{\hbar}{M} \\
&\times\big[\eta_{jn}\cos\left((n\!-\!m)\Omega t\right)+\delta_{jn}\sin\left((n\!-\!m)\Omega t\right)\big], \nonumber \\
G^j_{\gamma,-\mu}(t) = & \sum_{n,m}\frac{C_n^j C_m^j}{\nu_j}\frac{\hbar}{M} \\
&\times\big[\gamma_{jn}\cos\left((n\!-\!m)\Omega t\right)-\mu_{jn}\sin\left((n\!-\!m)\Omega t\right)\big], \nonumber \\
K^j(t) = & 4\hbar\kappa_j- 2 D^j_{\delta_,-\eta}(t),
\\
D^j_{\mu,\gamma}(t) = & \sum_{n,m}\frac{C_n^j C_m^j}{\nu_j}\left(\frac{\beta_j}{2}+m\right)\hbar\Omega \\
&\times\big[\mu_{jn}\cos\left((n\!-\!m)\Omega t\right)+\gamma_{jn}\sin\left((n\!-\!m)\Omega t\right)\big], \nonumber
\\
D^j_{\delta,-\eta}(t) = & \sum_{n,m}\frac{C_n^j C_m^j}{\nu_j}\left(\frac{\beta_j}{2}+m\right)\hbar\Omega \\
&\times\big[\delta_{jn}\cos\left((n\!-\!m)\Omega t\right)-\eta_{jn}\sin\left((n\!-\!m)\Omega t\right)\big]. \nonumber
\end{align}
Here, $G^j_{\eta,\delta}(t)$ plays the role of the time-dependent friction force, where sequences of constants $\eta_{jn}$ and $\delta_{jn}$ can be calculated for a given atom-ion potential, $\nu_j$ is the frequency of a reference oscillator{, $\beta_j/2$ denotes a characteristic exponent}, $C_n^j$, $C_m^j$ are the coefficients in a solution of Mathieu equation of an ion in a Paul trap without the buffer gas \cite{leibfried}, {and $\kappa_j$ is some coefficient that will be defined later separately for a condensed or a non-condensed reservoir.}

One can check that free terms $G^j_{\gamma,-\mu}(t)$ and $D^j_{\mu,\gamma}(t)$ assure that the ion energy cannot drop below the ground state energy of the secular trap even if the temperature of the atomic gas is lower. $\mu_{jn}$ and $\gamma_{jn}$ are sequences of constants depending on interaction potential. The form of Eq.~\eqref{liniowej} and Eq.~\eqref{kwadratowej} is general regardless of the interaction potential. They are valid both for an ultracold gas and a Bose-Einstein condensate, but the sequences of constants $\eta_{jn}$, $\delta_{jn}$, $\mu_{jn}$, $\gamma_{jn}$ and $\kappa_j$ have different functional forms {that will be introduced soon}.

{In order to derive their explicit form we use the following identity to calculate the integrals with respect to time variable}
\begin{equation}
\begin{split}
&\int_0^{\infty} d\tau e^{s_1 i \tau(\omega_{\mbf{k}}-\omega_{\mbf{k}'})+s_2 i \Omega((\beta_j/2+n)\tau-(n-m)t) }\\
&=e^{-i s_2 \Omega (n-m)t}\left(\pi \delta(s_1 (\omega_{\mbf{k}}-\omega_{\mbf{k}'})+s_2\Omega(\beta_j/2+n))\right.\\
&\left.+i\frac{{\cal{ P}}}{s_1 (\omega_{\mbf{k}}-\omega_{\mbf{k}'})+s_2\Omega(\beta_j/2+n)}\right),
\end{split}
\end{equation}
where $s_1$ and $s_2$ can be equal to $+$, $-$ or $0$ and $\delta(\dots)$ and $\cal{P}\dots$ denote Dirac delta and Principal value distributions. With help of above relation and Eq.~\eqref{2rzad} in case of a reservoir consisting of a non-condensed Bose gas {we obtain}
\begin{align}
\label{KappaGas}
\kappa_j = & \frac{1}{2\hbar^2}\sum_{\mbf{k},\mbf{k'}} |c_{\mbf{k},\mbf{k'}}|^2  (k_j-k'_j)^2 (\bar{n}_{\mbf{k'}}\bar{n}_{\mbf{k}}+\bar{n}_{\mbf{k}}) \frac{\cal{P}}{\omega_{\mbf{kk'}}}, \\
\eta_{jn} = & \frac{\pi}{2\hbar^2}\sum_{\mbf{k},\mbf{k'}} |c_{\mbf{k},\mbf{k'}}|^2 (k_j-k'_j)^2 (\bar{n}_{\mbf{k'}}\bar{n}_{\mbf{k}}+\bar{n}_{\mbf{k}})\\
& \times\big[\delta(\omega_{\mbf{kk'}}+\Omega(\beta_j/2+n))-\delta(\omega_{\mbf{kk'}}-\Omega(\beta_j/2+n))\big], \nonumber \\
\mu_{jn} = & \frac{\pi}{2\hbar^2}\sum_{\mbf{k},\mbf{k'}} |c_{\mbf{k},\mbf{k'}}|^2 (k_j-k'_j)^2 (\bar{n}_{\mbf{k'}}\bar{n}_{\mbf{k}}+\bar{n}_{\mbf{k}})\\
& \times\big[\delta(\omega_{\mbf{kk'}}+\Omega(\beta_j/2+n))+\delta(\omega_{\mbf{kk'}}-\Omega(\beta_j/2+n))\big], \nonumber \\
\delta_{jn} = & \frac{1}{2\hbar^2}\sum_{\mbf{k},\mbf{k'}} |c_{\mbf{k},\mbf{k'}}|^2 (k_j-k'_j)^2 (\bar{n}_{\mbf{k'}}\bar{n}_{\mbf{k}}+\bar{n}_{\mbf{k}})\\
& \times \left( \frac{\cal{P}}{\omega_{\mbf{kk'}}+\Omega(\beta_j/2+n)} +\frac{\cal{P}}{\omega_{\mbf{kk'}}-\Omega(\beta_j/2+n)}\right), \nonumber \\
\label{GammaGas}
\gamma_{jn} = & \frac{1}{2\hbar^2}\sum_{\mbf{k},\mbf{k'}} |c_{\mbf{k},\mbf{k'}}|^2 (k_j-k'_j)^2 (\bar{n}_{\mbf{k'}}\bar{n}_{\mbf{k}}+\bar{n}_{\mbf{k}})\\
& \times \left( \frac{\cal{P}}{\omega_{\mbf{kk'}}+\Omega(\beta_j/2+n)} -\frac{\cal{P}}{\omega_{\mbf{kk'}}-\Omega(\beta_j/2+n)}\right), \nonumber
\end{align}
where $\omega_{\mbf{kk'}}\equiv \omega_{\mbf{k}}-\omega_{\mbf{k'}}$. For a Bose-condensed reservoir
\begin{align}
\label{KappaBEC}
\kappa_j = & -\frac{\rho_0 L^3}{2\hbar^2}\sum_{\mbf{q}}(u_\mbf{q}+v_\mbf{q})^2 |c_{0,\mbf{q}}|^2  q_j^2 \frac{\cal{P}}{\omega_q}, \\
\eta_{jn} = & -\frac{\pi \rho_0 L^3}{2\hbar^2}\sum_{\mbf{q}}(u_\mbf{q}+v_\mbf{q})^2 |c_{0,\mbf{q}}|^2 q_j^2 \\
&\times\big[\delta(\omega_{\mbf{q}}+\Omega(\beta_j/2+n))-\delta(\omega_{\mbf{q}}-\Omega(\beta_j/2+n))\big], \nonumber \\
\mu_{jn} = & \frac{\pi \rho_0 L^3}{2\hbar^2}\sum_{\mbf{q}}(u_\mbf{q}+v_\mbf{q})^2 |c_{0,\mbf{q}}|^2 q_j^2 2\left(\bar{n}_{\mbf{q}}+\frac12\right)\\
&\times\big[\delta(\omega_{\mbf{q}}+\Omega(\beta_j/2+n))+\delta(\omega_{\mbf{q}}-\Omega(\beta_j/2+n))\big], \nonumber \\
\delta_{jn} = & -\frac{\rho_0 L^3}{2\hbar^2}\sum_{\mbf{q}}(u_\mbf{q}+v_\mbf{q})^2 |c_{0,\mbf{q}}|^2 q_j^2\\
&\times \left( \frac{\cal{P}}{\omega_{\mbf{q}}+\Omega(\beta_j/2+n)} +\frac{\cal{P}}{\omega_{\mbf{q}}-\Omega(\beta_j/2+n)}\right), \nonumber \\
\label{GammaBEC}
\gamma_{jn} = & \frac{\rho_0 L^3}{2 \hbar^2}\sum_{\mbf{q}}(u_\mbf{q}+v_\mbf{q})^2 |c_{0,\mbf{q}}|^2 q_j^2 2\left(\bar{n}_{\mbf{q}}+\frac12\right)\\
&\times \left( \frac{\cal{P}}{\omega_{\mbf{q}}+\Omega(\beta_j/2+n)} -\frac{\cal{P}}{\omega_{\mbf{q}}-\Omega(\beta_j/2+n)}\right). \nonumber
\end{align}

{In equations of motion \eqref{liniowe} and \eqref{kwadratowe} in the following text we omit} contributions from principal values present in $\kappa_j$, $\gamma_{jn}$ and $\delta_{jn}$ coefficients. One can check that their main role is renormalization of the trap $a_j$ and $q_j$ parameters due to the interactions with surrounding atomic gas. In typical experimental realizations, such effects are negligibly small. Moreover, we have verified that omitting terms containing principal values {do} not change the cooling rate and the final energies of the ion for small $a_j$ and $q_j$ parameters, relevant for current experiments.

{
With this simplifications equations of motion read
\begin{equation}
\label{liniowe}
\begin{split}
\dot{\overline{r}}_j&=\overline{p}_j/M\\
\dot{\overline{p}}_j&=- M\tilde{\omega}^2_j(t) \overline{r}_j -  \overline{p}_j \sum_{n,m}C^j_n C^j_m \tilde{\eta}_{jn}\cos\left((n\!-\!m)\Omega t\right),
\end{split}
\end{equation}
{where $\tilde{\eta}_{jn}=\frac{2 \hbar}{\nu_j M} \eta_{jn}$, $\tilde{\omega}_j^2(t) = \omega_j^2(t)-\sum_{n,m}C^j_n C^j_m \Omega (\beta_j/2+m)\tilde{\eta}_{jn}\sin\left((n\!-\!m)\Omega t\right)$.}
{In case of a Bose-Einstein condensate $\tilde{\eta}_{jn}$ can be calculated analytically
\begin{equation}
\begin{split}
\tilde{\eta}_{jn}=&\frac{4\sqrt{2}\pi}{3} \frac{m^{1/2}(m+M)^2}{M^{5/2}} \rho {a^j_{i}}^3\left(\frac{f(\tilde{q}_{jn})}{a^j_{i}}\right)^2\\
&\times \frac{\Omega^{3/2}}{\nu_j^{1/2}}|\beta_j/2+n|^{3/2}\mrm{sign}(\beta_j/2+n)
\end{split}
\end{equation}
where  $a^j_{i}=\sqrt{\hbar/(M\nu_j)}$, $\tilde{q}_{jn}=\sqrt{2 m \Omega |\beta_j/2+n|/\hbar}$.} The term $\sum_{n,m}C^j_n C^j_m \tilde{\eta}_{jn}\cos\left((n\!-\!m)\Omega t\right)$ in the above equations plays the role of the time-dependent friction force,}
$f(\tilde{q}_{jn})$ denotes the scattering amplitude for potential $V(r)$ calculated in the first-order Born approximation. Similar procedure may be applied to equations involving expectation values of operators quadratic in position and momentum, which yield
}

\begin{align}
\label{kwadratowe}
(\dot{\overline{r_j p_j}}+\dot{\overline{p_j r_j}})=&-2M\tilde{\omega}_j^2(t) \overline{r_j^2}+2\overline{p_j^2}/M \nonumber \\
-\sum_{n,m}&\left[ (\overline{r_j p_j}+\overline{p_j r_j}) \tilde{\eta}_{jn}\chi^j_{nm}(t)+2 \hbar \tilde{\mu}_{jn}\sigma^j_{nm}(t)\right],\nonumber\\
\dot{\overline{r_j^2}}=&\frac{1}{M} (\overline{r_j p_j}+\overline{p_j r_j}),\\
\dot{\overline{p_j^2}}=&-M\tilde{\omega}^2(t) (\overline{r_j p_j}+\overline{p_j r_j}) \nonumber \\
-2& \sum_{n,m} \left[ \overline{p_j^2}\tilde{\eta}_{jn}-\hbar \Omega M\!\! \left(\!\frac{\beta_j}{2}\!+\!m\!\right)\!\tilde{\mu}_{jn} \right] \chi^j_{nm}(t),\nonumber
\end{align}
{where  $\tilde{\mu}_{jn}=\frac{2 \hbar}{\nu_j M} \mu_{jn}$,} $\sigma^j_{nm}(t) = C^j_n C^j_m \sin((n-m)\Omega t)$ and $\chi^j_{nm}(t) = C^j_n C^j_m \cos((n-m)\Omega t)$. {Free term $\tilde{\mu}_{jn}$ assures that the ion energy cannot drop below the ground state energy of the secular trap even if the temperature of the atomic gas is lower. In case of a BEC it can be calculated analytically}
\begin{equation}
\begin{split}
\tilde{\mu}_{jn}=&\frac{2\sqrt{2}\pi}{3} \frac{m^{1/2}(m+M)^2}{M^{5/2}} \rho {a^j_{i}}^3\left(\frac{f(\tilde{q}_{jn})}{a^j_{i}}\right)^2\\
&\times \frac{\Omega^{3/2}}{\nu^{1/2}}|\beta/2+n|^{3/2}(2 \bar{n}_{\tilde{q}_n}+1).
\end{split}
\end{equation}



{ The form of the equations of motion is similar to classical equations describing particle in harmonic potential with time-dependent frequency and time-dependent damping. For small $a_j,q_j^2\ll 1$ (typical in experimental realizations) and sufficiently small gas density only limited number of terms in sums containing gas-dependent coefficients would be important from the point of view of the dynamics.}

\begin{figure}
\subfigure[]{\includegraphics[width=\linewidth,clip]{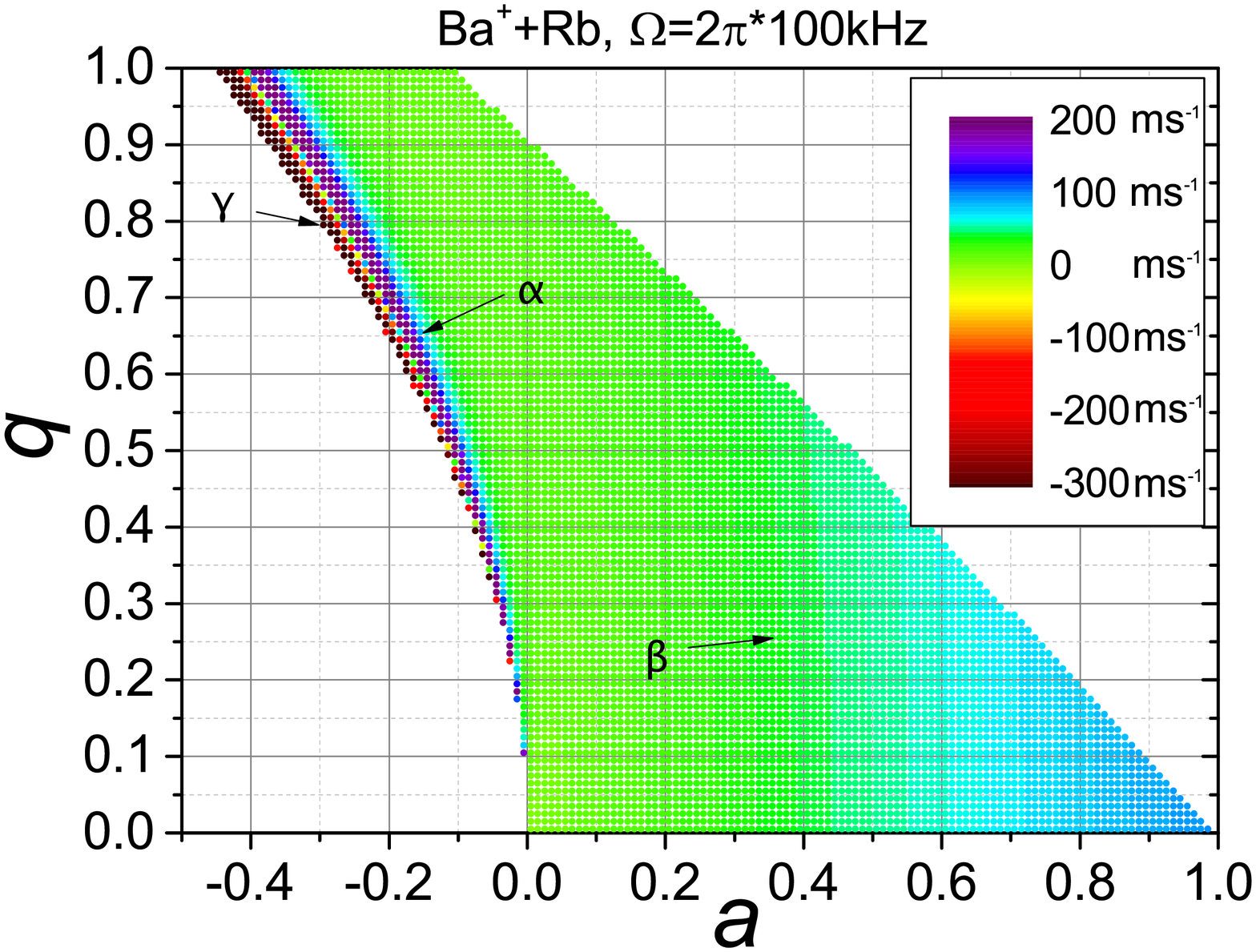}}\\
\subfigure[]{\includegraphics[width=\linewidth,clip]{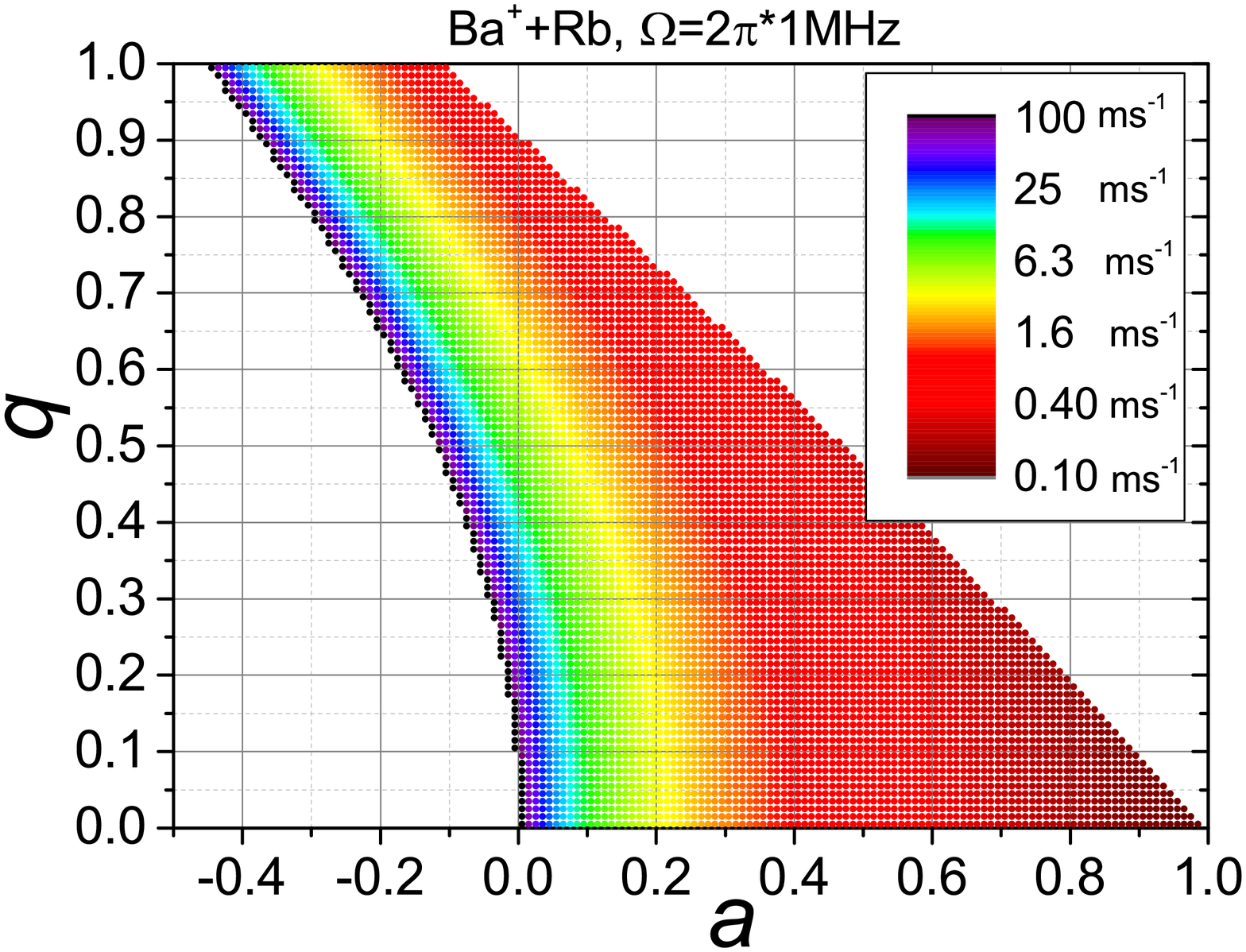}}\\
\subfigure[]{\includegraphics[width=\linewidth,clip]{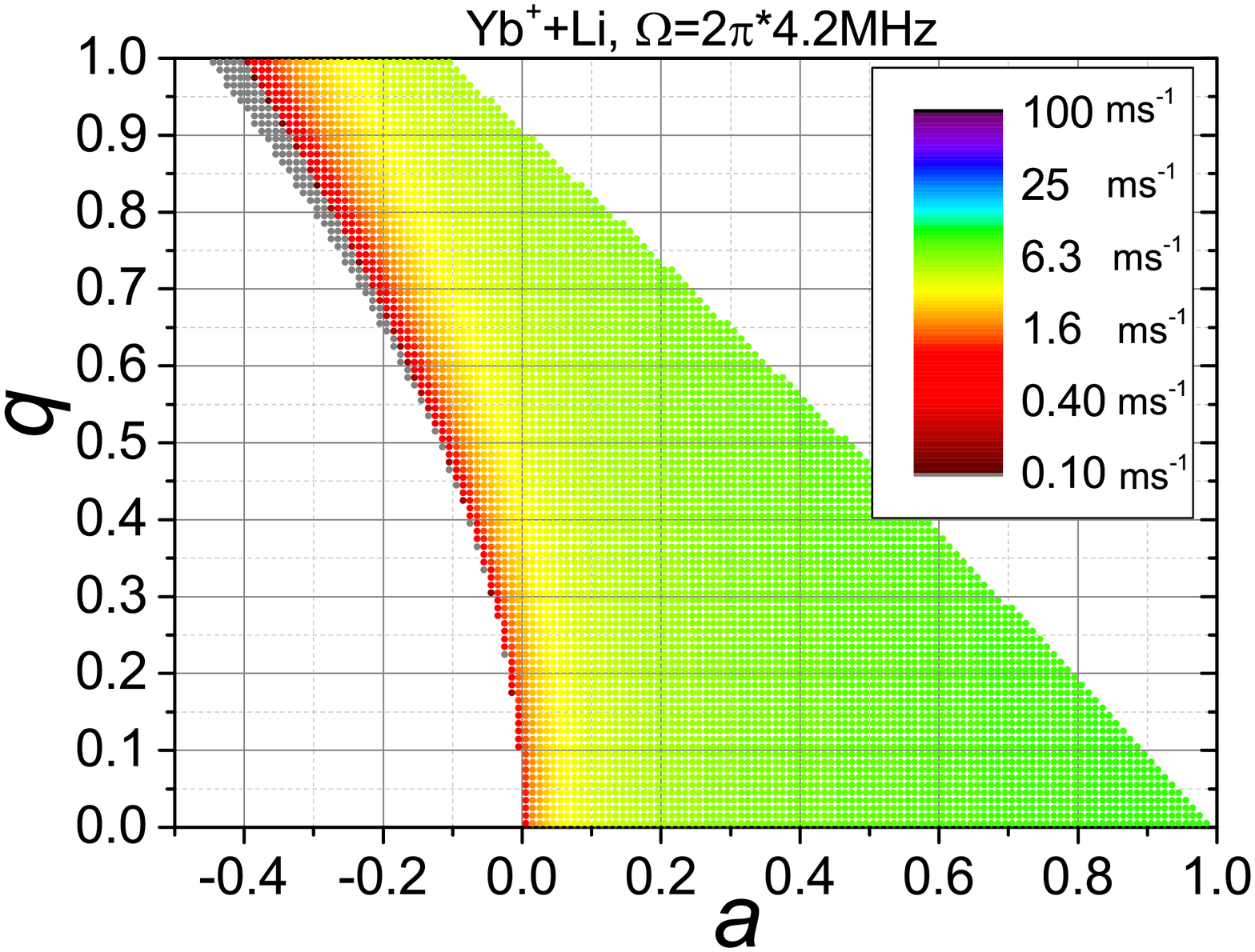}}\\
\caption{\label{Fig:bifurkacja}
(Color online). {Cooling rate
in one spatial direction for ${}^{138}$Ba${}^{+}$ ion ((a) and (b)) or ${}^{174}$Yb${}^{+}$ (c) immersed in an ultracold gas of ${}^{87}$Rb ((a) and (b)) or ${}^{7}$Li (c), at $T=200$~$n$K, scattering length $a_{sc}=R^{\star}$,  $\Omega=2\pi$~100kHz (a), $\Omega=2\pi$~1MHz (b) or $\Omega=2\pi$~4.2MHz (c),  $\rho_{GAS}=10^{13}/{\mrm{cm}}^3$ (a) or $\rho_{GAS}=10^{12}/{\mrm{cm}}^3$ ((b) and (c)). Points $\alpha$, $\beta$, $\gamma$ in the top panel represent three different regions of ion dynamics. (see text for details).}
}

\end{figure}

\begin{figure}\includegraphics[width=\linewidth,clip]{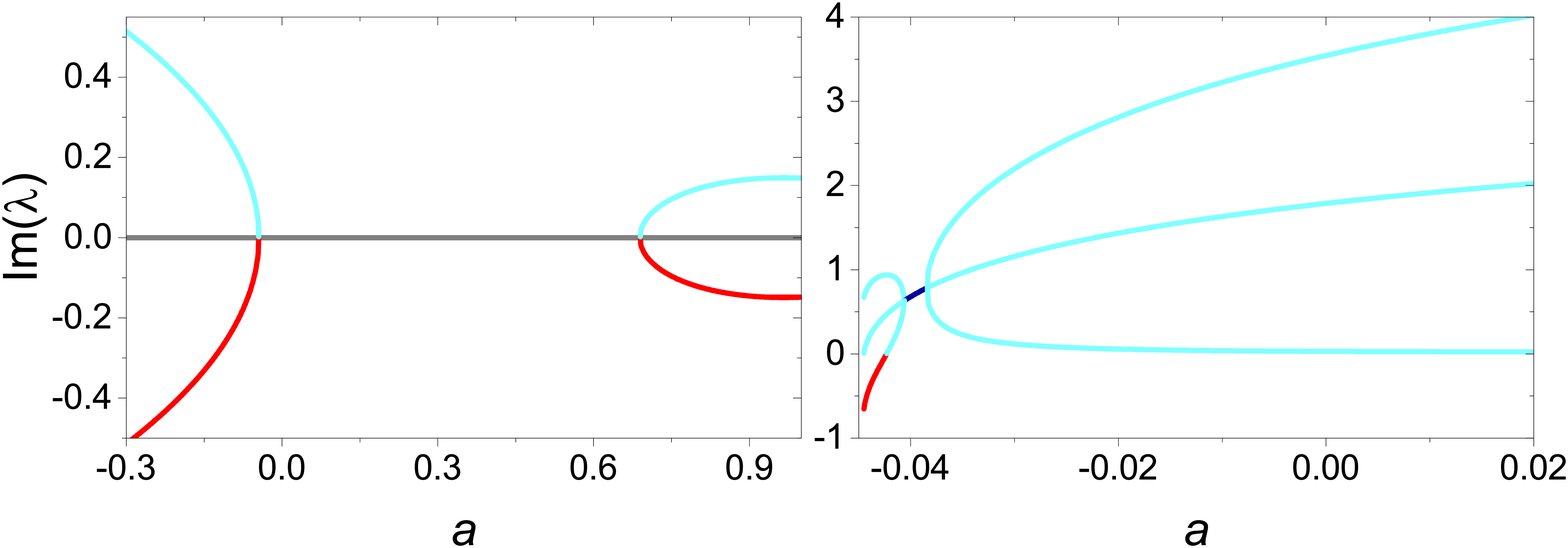}
\caption{\label{Fig:urojone}
(Color online). Imaginary part of the characteristic exponent for $q=0.3$ without (left) and with (right) the buffer gas (parameters as in the top panel of Fig.~\ref{Fig:bifurkacja}). {Grey line corresponds to stable solution for the Paul trap ($\Im (\lambda)=0$), dark blue line - the region of under-damped harmonic oscillator ($\Im (\lambda)>0$, $\Re(\lambda)\neq 0$),  light blue line - over-damped harmonic oscillator ($\Im (\lambda)>0$, $\Re(\lambda)=0$), red line - nonstable harmonic oscillator ($\Im (\lambda)<0$).}
}
\end{figure}

In order to solve Eq.~\eqref{liniowe} we use the following ansatz ${\mbf v}(t) = \sum_{n=-\infty}^{\infty} {\mbf w}_n e^{i \lambda \Omega t} e^{i n \Omega t}$ (similar to used in \cite{IdziaszekZoller2011,Nguyen2012,Zoller}), while for Eq.~\eqref{kwadratowe} we use ${\mbf v}(t) = \sum_{n=-\infty}^{\infty} {\mbf w}_n e^{i \lambda \Omega t} e^{i n \Omega t}+\sum_{n=-\infty}^\infty {\mbf u}_n e^{i n \Omega t}$ where ${\mbf v}(t)$ represents $(\overline{r}(t),\overline{p}(t))$ for Eq.~\eqref{liniowe} and
$(\overline{rp}(t)+\overline{pr}(t),\overline{r^2}(t),\overline{p^2}(t))$  for Eq.~\eqref{kwadratowe}, and $\lambda$ is a complex-valued characteristic exponent. Real part of $\lambda$ describes the secular frequency of oscillations of the ion in an effective trap and the imaginary part describes the cooling rate {(see Figs.~\ref{Fig:bifurkacja} and \ref{Fig:urojone})}. In case of cooling ($\Im(\lambda)>0$ - discussed below) the first part of the ansatz for Eq.~\eqref{kwadratowe} goes to zero for large times and the second one represents the asymptotic solution with $\mbf{u}_0$ being the average value of $\mbf{v}(t)$ with respect to {the time scale given by the RF frequency $\Omega$.}

{\section{VII. Cooling rates}} Let us analyze the ion dynamics for some typical experimental parameters. Fig.~\ref{Fig:bifurkacja} shows the cooling rates of a Ba$^{+}$ ion immersed in an ultracold gas of Rb atoms for parameters similar to used in the experiment \cite{Schmid2010} (middle panel) and for a weaker trap and a higher gas density (top panel). {The bottom panel of Fig.~\ref{Fig:bifurkacja} presents the cooling rates of Yb$^{+}$ ion immersed in an ultracold gas of Li atoms.} The white color represents the unstable regions of the Paul trap without the buffer gas. In general there are three different regimes of ion dynamics immersed in an ultracold gas: (i) under-damped, (ii) over-damped harmonic motion, (iii) heating. They are marked as points $\alpha$, $\beta$ and $\gamma$, respectively in the top panel of Fig.~\ref{Fig:bifurkacja}. In regime (ii) the ion motion is similar to over-damped harmonic oscillator. The real part of the characteristic exponent $\lambda$ drops to zero and the energy asymptotically reaches its final value. However, there still exists the micromotion, that adds the periodic modulation of the energy. For parameters of Fig.~\ref{Fig:bifurkacja}(a) the region of over-damped motion dominates, but this is only due to relatively weak trapping and high density of the atomic gas. In typical experimental realizations, however, the stability region corresponds mainly to the cooling behavior of the under-damped motion, see Fig.~\ref{Fig:bifurkacja}(b).

\begin{figure}\includegraphics[width=\linewidth,clip]{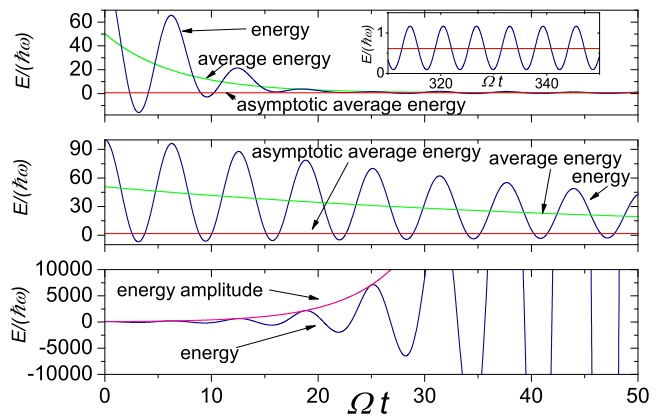}
\caption{\label{Fig:energie}
{(color online). Energy of the ion (expectation value of $H_S$, blue line), average energy with respect to the period of the radio frequency modulation (green), asymptotic average energy (red), and energy amplitude (pink) for parameters of point $\alpha$ (top panel), $\beta$ (middle panel), and $\gamma$ (bottom panel) marked in Fig.~\ref{Fig:bifurkacja}.}
}
\end{figure}

{
The time dependence of the energy for three described regimes is shown in Fig.~\ref{Fig:energie}, for parameters corresponding to points $\alpha$, $\beta$ and $\gamma$, respectively (cf. Fig.~\ref{Fig:bifurkacja} (a)).
In the regime represented by point $\alpha$ (upper panel) the ion energy decreases exponentially in time while the secular frequency of the trap is renormalized in the presence of cold reservoir. The inset shows the asymptotic behaviour at large $t$. Since the energy is not conserved in the presence of time-dependent potential, one can observe remaining oscillations around the final value. For point $\gamma$ (bottom panel) the motion of the ion is unstable - it gains energy exponentially from the time-dependent trap. The pink line depicts the growing amplitude of the oscillatory motion. For parameters of point $\beta$ (middle panel) one can observes the net cooling effect (green line), after averaging out over fast micromotion. However, the time-scale of cooling in this regime is much longer than for cooling in region represented by point $\alpha$.}

Fig.~\ref{Fig:urojone} shows the imaginary part of the characteristic exponent $\lambda$ for an isolated ion (left panel) and for the ion immersed in cold reservoir (right panel) for parameters of Fig.~\ref{Fig:bifurkacja}(a). For values larger than zero the ion motion is being damped (regime (i) or (ii)), while for negative values the motion is unstable (regime (iii)). This figure explains the presence of three different regions of ion dynamics. In the absence of the buffer gas there are only two regions - stable and unstable and there is no cooling inside the stable region, because the imaginary part of the characteristic exponent is zero there. Introduction of the buffer gas increases
$\Im(\lambda)$ and in the comparison to {the isolated ion case a new region appears (all solutions marked in blue) where the ion motion is damped, and its energy decreases.}

There is a possibility of creating molecular states of an atom and an ion in the course of collisions. Such states have been predicted to emerge in a classical simulation of ion-atom collisions \cite{Vuletic2012} in the presence of the time-dependent RF potential. Ref.~\cite{Vuletic2012} shows that during the collision the particles can be bound for relatively long time, and the work performed by the time-dependent electric field constitutes a significant source of heating, shifting the final temperatures of ions in the sympathetic cooling up to mK regime. Our formalism neglects the effect of bound states association in the atom-ion collisions, but our analysis shows that this process should not be significant in the low-energy quantum regime, because the probability of creation of a molecular complex will be significant only when the resonance condition is fulfilled. In order to verify this assumption we estimated the probability of transition to molecular complex during a single collision using time-dependent perturbation theory. For Rb-Yb$^+$  collisions and $a_{sc} = R^\ast = 307$nm the probability of association of a molecular complex in a single collision is $P = 0.007$, while for $a_{sc} = - R^\ast = - 307$nm is larger: $P = 0.127$. In the association process one or more energy quanta are transferred from the collision complex to RF field, and the probability of association strongly depends on the fulfillment of the resonance condition between the initial and final molecular states. Therefore, by appropriate selection of the trap parameters it should be possible to detune from the resonance, and finally reduce the probability of association.


\section{VIII. Concluding Remarks}

{
{Based on the theory of quantum stochastic processes we have developed a master equation for the system in the time-dependent external potential, in which a single trapped ion is brought into a contact with an} ultracold gas in a condensed or an non-condensed phase. We have investigated three different stability regimes of the ion motion. Furthermore, we have studied experimentally relevant sets of parameters and we have {calculated} cooling rates for Ba$^+$ ion immersed in a {Rb} atoms and Yb$^+$ ion immersed in a Li reservoir. {In typical experimental realizations also so called excess micromotion constitutes an additional source of heating. We plan to investigate this issue in future research.}
}


\section{Acknowledgments}
We thank K. Jachymski, R. Gerritsma and E. Hudson for stimulating discussions and C. Sias, A. H\"{a}rter and A. Kr\"{u}kow for providing us with experimental parameters. This work was supported by the Foundation for Polish Science International PhD Projects Programme co-financed by the EU European Regional Development Fund and National Science Centre {(ZI) (Grants No. DEC-2011/01/B/ST2/02030 (ZI) and DEC-2012/07/N/ST2/02879 (MK)).}

\bibliography{master}

\begin{appendix}

\section{Appendix: Probability of associating bound states during the collision process}\label{App:BoundStates}
In order to estimate the probability of transition to molecular states during a single atom-ion collision, we have developed a one-dimensional model assuming that atom approaches the ion along one-dimensional trajectory, which should be a good approximation to real three-dimensional scattering process in a time-dependent field \cite{Vuletic2012}. First, we have transformed the total Hamiltonian with the help of the Cook, Shankland, Wells transformation \cite{Nguyen2012}, separating it into static and time-dependent parts:
\begin{align}
H(t) = & H_{0} + \tilde{H}(t) \\
H_0 = & \frac{\hat{p}_i^2}{2 M} + \frac{M}{2} \nu^2 x^2+ \frac{\hat{p}_a^2}{2 m} + V(|x-x_a|) \\
\tilde{H}(t) = & - M (\gamma \nu)^2 x^2 \cos \left(2 \Omega t\right) \\
& + 2 i \hbar \gamma \nu \left(
x \frac{\partial}{\partial x} + \frac12 \right) \sin \left(\Omega t \right)
\end{align}
The first part $H_{0}$ contains kinetic energies of the atom and ion, the static part of the Paul trap with $\nu$ denoting the frequency of the reference harmonic oscillator \cite{leibfried}, and the atom-ion interaction is given by $V(|x|)$. The second part $\tilde{H}(t)$ contains two time-dependent terms oscillating with frequency of the RF field $\Omega$ and $2 \Omega$, respectively. The coefficient $\gamma$
\begin{equation}
\gamma=\frac{1}{\sqrt{2(1+\frac{2a}{q^2})}}.
\end{equation}
depends on the ratio of the static and dynamic amplitudes of the RF field.

We calculate the transition probability from the scattering state to the bound state of $H_0$ in the first-order time-dependent perturbation theory, treating $\tilde{H}(t)$ as perturbation \cite{Nguyen2012}. The initial and final states of $H_0$ can be represented as
\begin{equation}
\Psi(x,x_a) = \sum_{n=0}^{\infty} \phi_n(x) \psi_n(x_a)
\end{equation}
where $\phi_n(x)$ are the wave functions of the ion in the static trap
\begin{equation}
\left(\frac{\hat{p}^2}{2 M} + \frac12 M \nu^2 x^2 \right) \phi_n(x) = E_n \phi_n(x)
\end{equation}
and $\psi_n(x_a)$ are corresponding wave functions of the atom. We assume that initially the ion is in the ground-state of the Paul trap and the asymptotic kinetic energy of the free atom $E_a= (\hbar^2 k^2)/(2 m) < \hbar \nu$. In this case wave function of the atom for the channel $n=0$ takes the following asymptotic form at large distances:
\begin{equation}
\psi_0(x_a) \stackrel{|x_a| \to \infty}{\longrightarrow} e^{i k x_a} + f_{+} e^{i k |x_a|} + f_{-} \textrm{sgn}(x_a) e^{i k |x_a|}
\end{equation}
Here, $f_{+}$ and $f_{-}$ denote the scattering amplitudes corresponding to even and odd scattered waves, respectively.

The transition probability in the first-order perturbation theory is given by Fermi's golden rule. We sum independent contributions due to the transitions induced by the first and the second term in $\tilde{H}(t)$. Moreover, we include only the processes in which the energy quanta are transferred from the collision complex to the RF field, which lead to creation of molecular states. Hence the total probability of bound-state association in the single collision can be approximated by
\begin{equation}
P \approx P_{1} + P_{2},
\end{equation}
where
\begin{align}
P_{1} & = \frac{2 \pi}{\hbar} \left|\langle\Psi_i|\tilde{H}_1|\Psi_f\rangle\right|^2 \rho(E-2 \hbar \Omega) j^{-1} \\
P_{2} & = \frac{2 \pi}{\hbar} \left|\langle\Psi_i|\tilde{H}_2|\Psi_f\rangle\right|^2 \rho(E-\hbar \Omega) j^{-1} ,
\end{align}
with
\begin{align}
\tilde{H}_1 & = -  \frac12 M (\gamma \nu)^2 x^2\\
\tilde{H}_2 & = \hbar \gamma \nu \left(x \frac{\partial}{\partial x} + \frac12 \right)
\end{align}
Here, $E = \hbar^2 k^2/(2 m) + \frac12 \hbar \nu$ denotes the energy of the initial scattering state $\Psi_i$, while the energy of the final state $\Psi_f$ is $E-2\hbar \nu$ and $E-\hbar \nu$ for transitions induced by $\tilde{H}_1$ and $\tilde{H}_2$ terms, respectively, $j = \hbar k/m$ is the probability flux of the atoms scattering on the ion, and $\rho(E)$ is the density of states. The density of states for energy corresponding to the resonance was determined from the numerical energy spectrum calculated by the diagonalization of the Hamiltonian $H_0$.

The initial and final states were determined by numerically diagonalizing Hamiltonian $H_0$ in a box of size much larger than $R^\ast$. The secular frequency was $\nu = 2 \pi \times 1$MHz. We calculated the probability of transition to bound state for Rb-Yb$^{+}$ system, for two different scattering lengths $a_{sc} = R^\ast = 307$nm and $a_{sc} = - R^\ast = - 307$nm. The RF frequency in both cases was about $2 \pi \times 10$MHz, and its exact value was chosen in order to be resonant with a molecular level resonant with the transition induced by $\tilde{H}_2(t)$ (dominating perturbation). This should overestimate the probability of creation of a molecular state and in real experimental conditions one can always try to modify the RF frequency in order to detune from the resonance.

\end{appendix}

\end{document}